\documentclass[a4paper,11pt]{article}
\usepackage{pos}

\title{Search for dark matter annihilation signals from the Galactic Center with the H.E.S.S. Inner Galaxy Survey}

 \ShortTitle{Search for dark matter with the H.E.S.S. IGS dataset}

\author*[a]{Alessandro Montanari}
\author[a]{Emmanuel Moulin}
\author[b]{Denys Malyshev}

\affiliation[a]{IRFU, CEA, Universit\'e Paris-Saclay, F-91191 Gif-sur-Yvette, France}
\affiliation[b]{Institut f\"ur Astronomie und Astrophysik, Universit\"at T\"ubingen, Sand 1, D 72076 T\"ubingen, Germany}

\forColl{H.E.S.S.} 
\emailAdd{alessandro.montanari@cea.fr}
\emailAdd{emmanuel.moulin@cea.fr}
\emailAdd{denys.malyshev@astro.uni-tuebingen.de}

\abstract{The presence of dark matter (DM) is suggested by a wealth of astrophysical and cosmological measurements. However, its underlying nature is yet unknown. Among the most promising candidates are weakly interacting massive particles (WIMPs): particles with mass and coupling strength at the electroweak scale and thermally-produced in the early universe have a present relic density consistent with that observed today. WIMP self-annihilation would produce Standard Model particles including gamma-rays, which have been long-time recognized as a prime messenger to indirectly detect dark matter signals. The centre of the Milky Way is predicted as the brightest source of DM annihilations. The H.E.S.S. collaboration is currently performing a survey of the inner region of the Milky Way, the Inner Galaxy Survey (IGS), intended to achieve the best sensitivity to faint and diffuse emissions in a region of several degrees around the Galactic Centre. We analyzed 2014-2020 observations taken with the five-telescope array to search for a DM annihilation signal. With the current dataset of about 550 hours, we found no significant excess and therefore derived strong constraints on the velocity-weighted annihilation cross-section. TeV thermal WIMPs can be probed in different annihilation channels.}

\FullConference{37$^{\rm{th}}$ International Cosmic Ray Conference (ICRC 2021)\\
		July 12th -- 23rd, 2021\\
		Online -- Berlin, Germany}

\begin{document}
\maketitle
\section{Dark Matter search with gamma-ray observations}
\label{sec:dmsearch}
There are many astrophysical and cosmological measurements that sustain the hypothesis of 85\% of the total matter content of the Universe being made of dark and non-baryonic particles~\cite{Adam:2015rua}. Many beyond-the-Standard-Model theories have tried to explain the nature of this so-called dark matter (DM). Weakly interacting massive particles (WIMPs), that have mass and coupling strength at the electroweak scale, constitute one of the most compelling class of stable DM candidates~\cite{Bertone:2010zza,Feng:2010gw,Roszkowski:2017nbc}. All the DM in the Universe can be assumed as made of WIMPs, if the latter are considered as thermally produced in the early Universe~\cite{Jungman:1995df}. Many experiments have been motivated with the aim to search for WIMPs and probe their non-gravitational properties, such as the direct detection of the WIMP-nucleus scattering  process~\cite{Schumann:2019eaa}, the production of WIMPs at particle colliders~\cite{Kahlhoefer:2017dnp} and the detection of the products of WIMPs decay and annihilation~\cite{Strigari:2018utn}. Dense astrophysical environments could host self-annihilation of WIMPs. From the annihilation process, gamma-rays could be produced in the final state from hadronization, radiation and decay of standard model particles. When the WIMP mass is high enough, these gamma-rays could be detected from the H.E.S.S. array of five Imaging Atmospheric Cherenkov Telescopes (IACTs). The differential flux of gamma-rays, from the self-annihilation of Majorana WIMPs, in a solid angle $\Delta\Omega$, is expressed as:
\begin{multline}
\label{eq:dmflux}
\frac{\rm d \Phi_\gamma}{\rm d E_\gamma} (E_\gamma,\Delta\Omega)=
\frac{\langle \sigma v \rangle}{8\pi m_{\rm DM}^2}\sum_f  BR_f \frac{\rm d N^f}{\rm d E_\gamma}(E_\gamma) \, J(\Delta\Omega) \\
\quad {\rm with} \quad  J(\Delta\Omega) =  \int_{\Delta\Omega} \int_{\rm los} \rho^2(s(r,\theta)) ds\, d\Omega \,.
\end{multline}
$\langle \sigma v \rangle$ is the velocity-weighted annihilation cross section averaged over the velocity distribution. $dN^f/dE_\gamma$ and $BR_{f}$ are the differential gamma-ray yield per annihilation in the channel $f$ and its branching ratio, respectively. 
the J-factor $J(\Delta\Omega)$ expresses the integral of the square of the DM density $\rho$ over the line of sight (los) $s$ and the solid angle $\Delta\Omega$. Since the DM density $\rho$ is assumed spherically symmetric, it depends only on the radial coordinate $r$ from the center of the DM halo. It writes as $r = \big(s^2 +r_{\odot}^2-2\,r_{\odot}\,s\, \cos\theta \big)^{1/2}$. The angle between the direction of observation and the Galactic Center (GC) is $\theta$. $r_\odot$, the distance of the observer to the GC, is assumed to be $r_\odot$ = 8.5 kpc~\cite{Ghez:2008ms}. Assuming a DM distribution to follow a cuspy DM profiles, conveniently described by the NFW~\cite{Navarro:1996gj} and Einasto~\cite{Springel:2008by} parametrizations, the centre of the Milky Way is predicted as the brightest source of DM annihilation. 
In the TeV mass range, the strongest constraints on WIMPs are given so far by 254 hours of observations of the GC region with H.E.S.S.~\cite{Abdallah:2016ygi}. Other compelling and complementary targets for the search of DM annihilation signals that have been recently observed by H.E.S.S. are unidentified {\it Fermi}-LAT objects (UFOs)~\cite{Abdallah:2021czg} as source candidates for DM subhalos populating the main Galactic Halo, ultra-faint dwarf galaxy satellites of the Milky Way recently detected by the Dark Energy Survey (DES)~\cite{Abdallah:2020sas}, and the dwarf irregular galaxy WLM~\cite{Abdallah:2021kzs}. We report here on the first results of a new search for DM annihilation signals from the GC region. To perform this search, the unprecedented dataset of very-high-energy (VHE, E $\gtrsim$ 100 GeV) observations of the GC under the Inner Galaxy Survey programme with the five-telescope H.E.S.S. array is used.

\section{H.E.S.S. observations and data analysis}
\label{sec:analysis}
\begin{figure*}[!ht]
\centering
\includegraphics[width=0.5\textwidth]{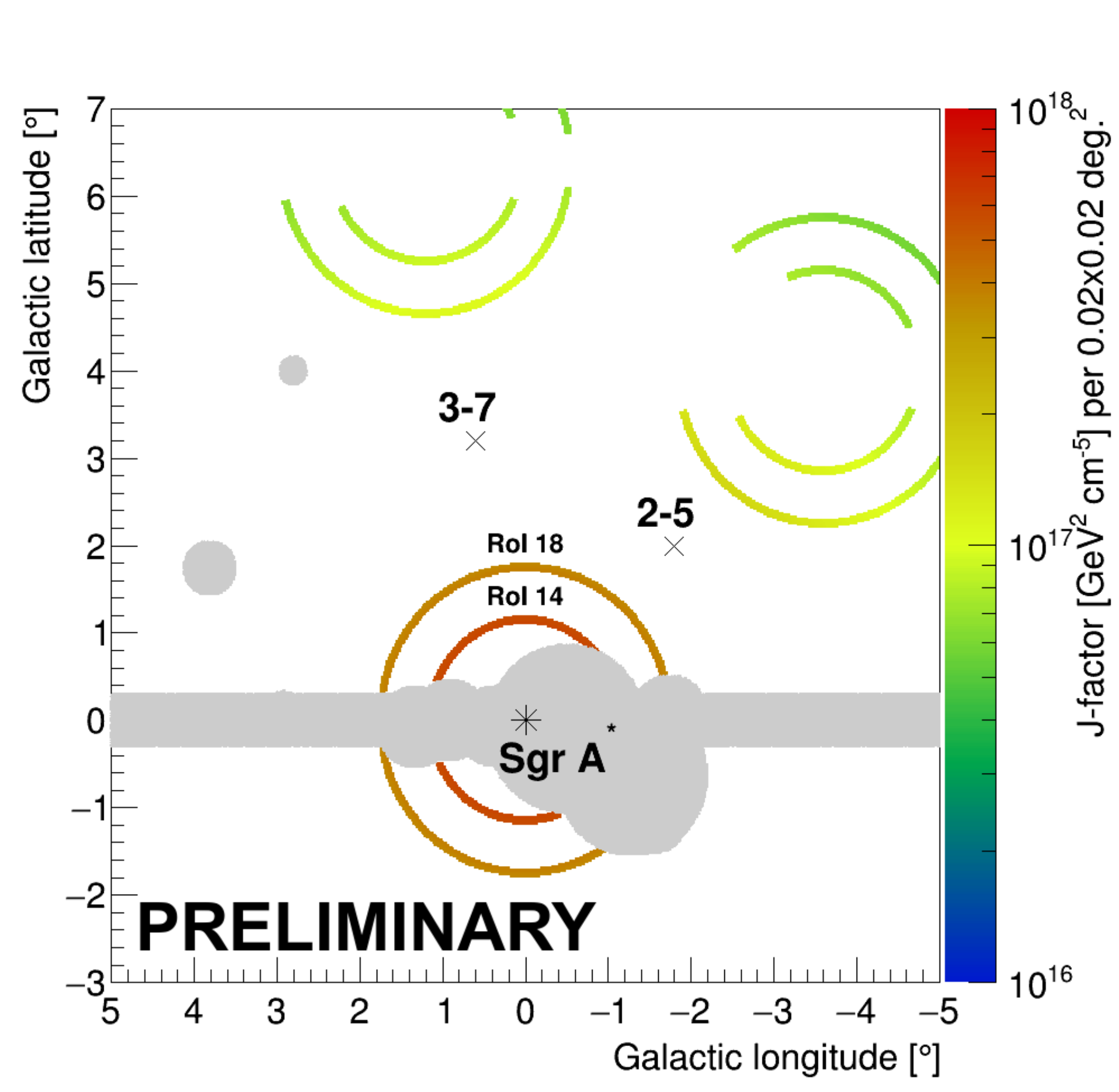}
\caption{An example of the reflected background method for background determination. Two pointing position of the IGS program are shown as black crosses. J-factor values integrated for the Einasto profile in pixels of 0.02$^\circ \times$0.02$^\circ$ size are displayed for ROI 14 and 18 as well as those obtained in the corresponding OFF regions. The grey-shaded area represents the set of masks used in this analysis to avoid contamination in the ROI from VHE sources. To keep the same solid angle size and acceptance in the ON and OFF regions, a similar exclusion of the masked regions is performed in each. The position of the supermassive black hole Sagittarius A* is marked with the black star.}
\label{fig:reflected_background}
\end{figure*}

Due to the location of the the H.E.S.S. instrument near the tropic of Capricorn,the central region of the Milky Way can be observed by H.E.S.S. under very good conditions. With the coverage of the inner several hundred parsecs around the GC region, the best sensitivity possible to DM annihilation signals and Galactic Center outflows is intended to be achieved with the IGS. The latter is the first-ever conducted VHE gamma-ray survey of the GC region. To cover regions not explored so far in VHE gamma rays, the telescope pointing positions of the IGS are chosen up to Galactic latitudes $b$ = +3.2$^\circ$. The dataset used here amounts to a total of 546 hours (live time) of high-quality data selected with the standard procedure~\cite{Aharonian:2006pe}, composed by 28-min data taking runs observed between 2014 and 2020. The zenith angles of the observations are kept lower than 40$^\circ$ in order to minimize the systematic uncertainties that could arise in the reconstruction of the event property. The present dataset averaged zenith angle is 18$^\circ$. At least then hours of acceptance-corrected time exposure are reached up to $b \approx$ +7$^\circ$. To select gamma-ray like events, a semi-analytical shower model technique based on the comparison of gamma-ray images in the camera from real events and predicted one from the model~\cite{2009APh32231D} is performed. The achieved angular resolution and energy resolution above 200 GeV are 0.06$^\circ$ (68\% containment radius) and 10\%, respectively. The region of interest (ROI) for the search of the DM annihilation signals is defined as a set of 25 rings, hereafter referred as to the ON region, centered on the nominal GC position, with inner radii from 0.5$^\circ$ to 2.9$^\circ$, and width of 0.1$^\circ$ each. Numerous VHE gamma-ray emitters populate the central region of the Milky Way~\cite{Aharonian:2009zk,Abramowski:2016mir,H.E.S.S.:2018zkf} and can contaminate the signal in the ROI. To avoid this, a conservative set of masks is used. The latter is shown in Fig.~\ref{fig:reflected_background}. For the measurement of the residual gamma-ray background on a run-by-run basis, a region taken symmetrically to the ON region from the pointing position for each ROI is defined. This region is hereafter referred as to the OFF region. The masks are applied similarly in the ON and OFF regions, to keep the same solid angle and acceptance. ON and OFF regions are sufficiently far away one from the other such that a significant gradient in the expected DM signal is obtained. Two examples of the reflected OFF regions for two ROI are shown in Fig.~\ref{fig:reflected_background}. A gradient in the residual background rate is expected across the telescope field of view, due to the different zenith angle of the observations. A difference in the zenith angles of the events is obtained between the ON and the OFF regions. For the considered ROI and pointing positions of the IGS, the difference of the means of the distributions of the ON and OFF event zenith angles is up to 1$^{\circ}$. A dependence of the gamma-ray-like background rate of 1\%-per-degree of difference in the zenith angle of the events is expected. A systematic uncertainty of 1\% in the normalisation of the measured event count distribution is therefore estimated and used in the analysis. 

The statistical data analysis is based on a 2-dimensional log-likelihood ratio test statistic (TS). To exploit expected spectral and spatial DM signal features against background, the TS is built on 67 logarithmically-spaced energy bins and 25 spatial bins corresponding to the ROI. For a given DM mass, the likelihood function writes:
\begin{multline}
\mathcal{L}_{\rm ij}({\bf N}^{\rm S},{\bf N}^{\rm B}|{\bf N}_{\rm ON},{\bf N}_{\rm OFF},{\bf \alpha}) = \frac{[\beta_{\rm ij}(N_{\rm ij}^{\rm S}+N_{\rm ij}^{\rm B})]^{N_{{\rm ON, ij}}}}{N_{{\rm ON, ij}}!}e^{-\beta_{\rm ij}(N_{\rm ij}^{\rm S}+ N_{\rm ij}^{\rm B})} \\
\frac{[\beta_{\rm ij}(N_{\rm ij}^{\rm S'}+\alpha_{\rm j}N_{\rm ij}^{\rm B})]^{N_{{\rm OFF, ij}}}}{N_{{\rm OFF, ij}}!} e^{-\beta_{\rm ij}(N_{\rm ij}^{\rm S'}+\alpha_{\rm j} N_{\rm ij}^{\rm B})} e^{-\frac{(1-\beta_{\rm ij})^2}{2\sigma^2_{\beta_{\rm ij}}}} \, .
\label{eq:lik}
\end{multline}
The number of measured events in the ON and OFF regions are expressed by $N_{\rm ON,ij}$ and $N_{\rm OFF,ij}$. The spectral and spatial bins are remarked by the $i$ and $j$ indices, respectively. The expected total number of DM events for the $(i,j)$ bin is $N_{\rm ij}^S$. To obtain it, the expected DM flux given in Eq.~(\ref{eq:dmflux}) is folded by the energy-dependent acceptance and energy resolution. $dN^f/dE_\gamma$ is computed with Pythia including final state radiative corrections~\cite{Cirelli:2010xx}. The expected total number of events in the bin $(i,j)$ is given by $N^S_{\rm ij} + N^B_{\rm ij}$. $\alpha_j = \Delta\Omega_{\rm OFF}/\Delta\Omega_{\rm ON}$ is the ratio between the angular size of the ON and OFF $j$ regions. This ratio is fixed to $\alpha_j \equiv 1$ since each OFF region $j$ is taken symmetrically to the ON region from the pointing position. To account for the systematic uncertainty on the normalisation of the measured energy count distribution, the likelihood function is completed with a Gaussian nuisance parameter. $\beta_{\rm ij}$ acts as a normalisation factor and $\sigma_{\beta_{\rm ij}}$ is the width of the Gaussian function (see, for instance, Refs.~\cite{Silverwood:2014yza,Lefranc:2015pza,Moulin:2019oyc}). To find $\beta_{\rm ij}$, a conditional maximization of the likelihood function $\rm d \mathcal{L}_{\rm ij}/d \beta_{\rm ij} \equiv 0$ is performed. In this analysis, $\sigma_{\beta_{\rm ij}}$ is fixed to 1\%. Constraints on $\langle \sigma v \rangle$ as a function of the DM mass $m_{\rm DM}$ are computed with the log-likelihood ratio TS as described in Ref.~\cite{Cowan:2010js}, when no significant excess is found and a positive searched signal $\langle \sigma v \rangle >$ 0 is assumed. When the TS value is higher than 2.71, the corresponding $\langle \sigma v \rangle$ is excluded at the 95\% confidence level (C.L.).

\section{Results and Conclusion}
\label{sec:results}
\begin{figure*}[ht!]
\includegraphics[width=0.5\textwidth]{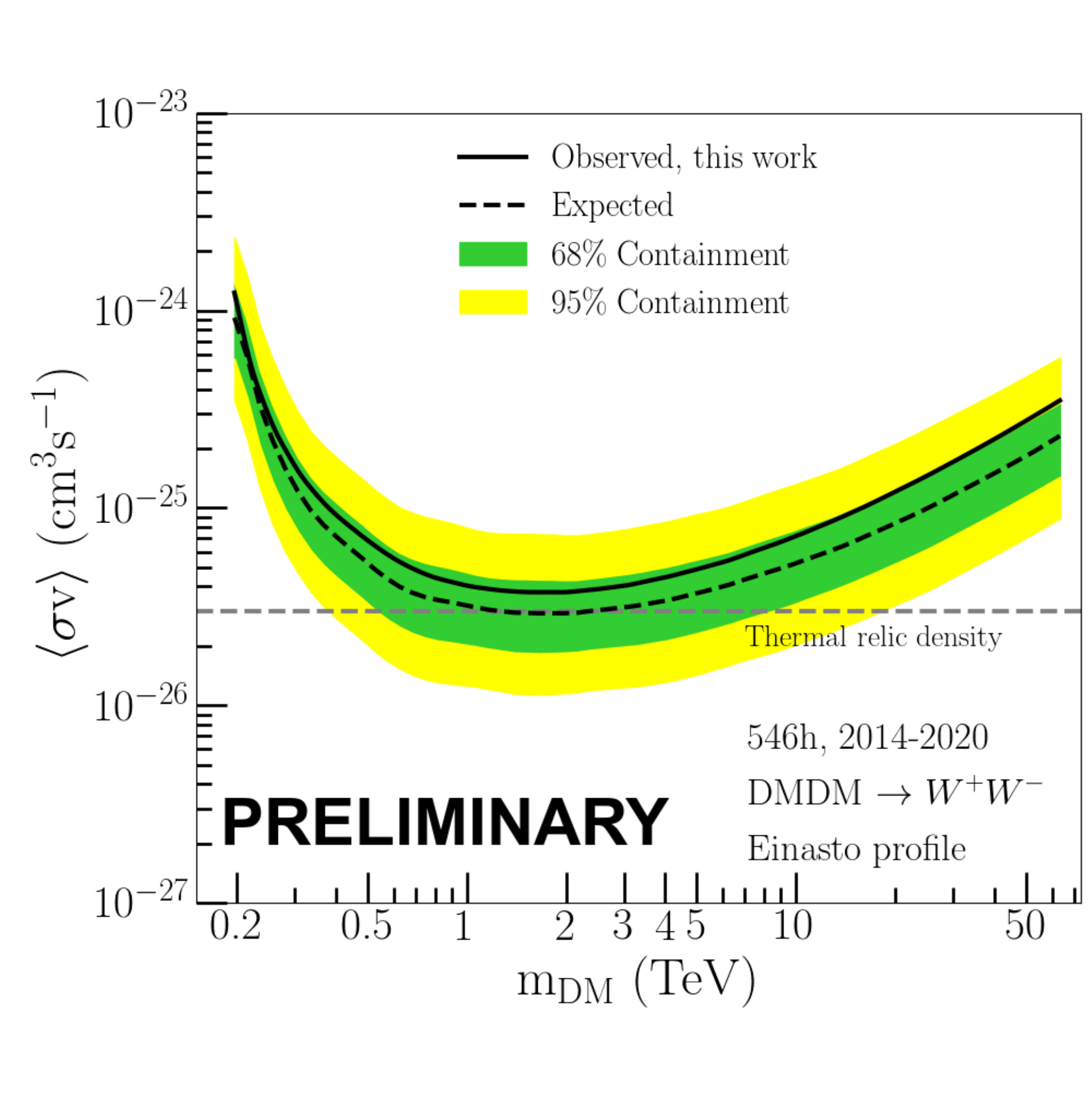}
\includegraphics[width=0.5\textwidth]{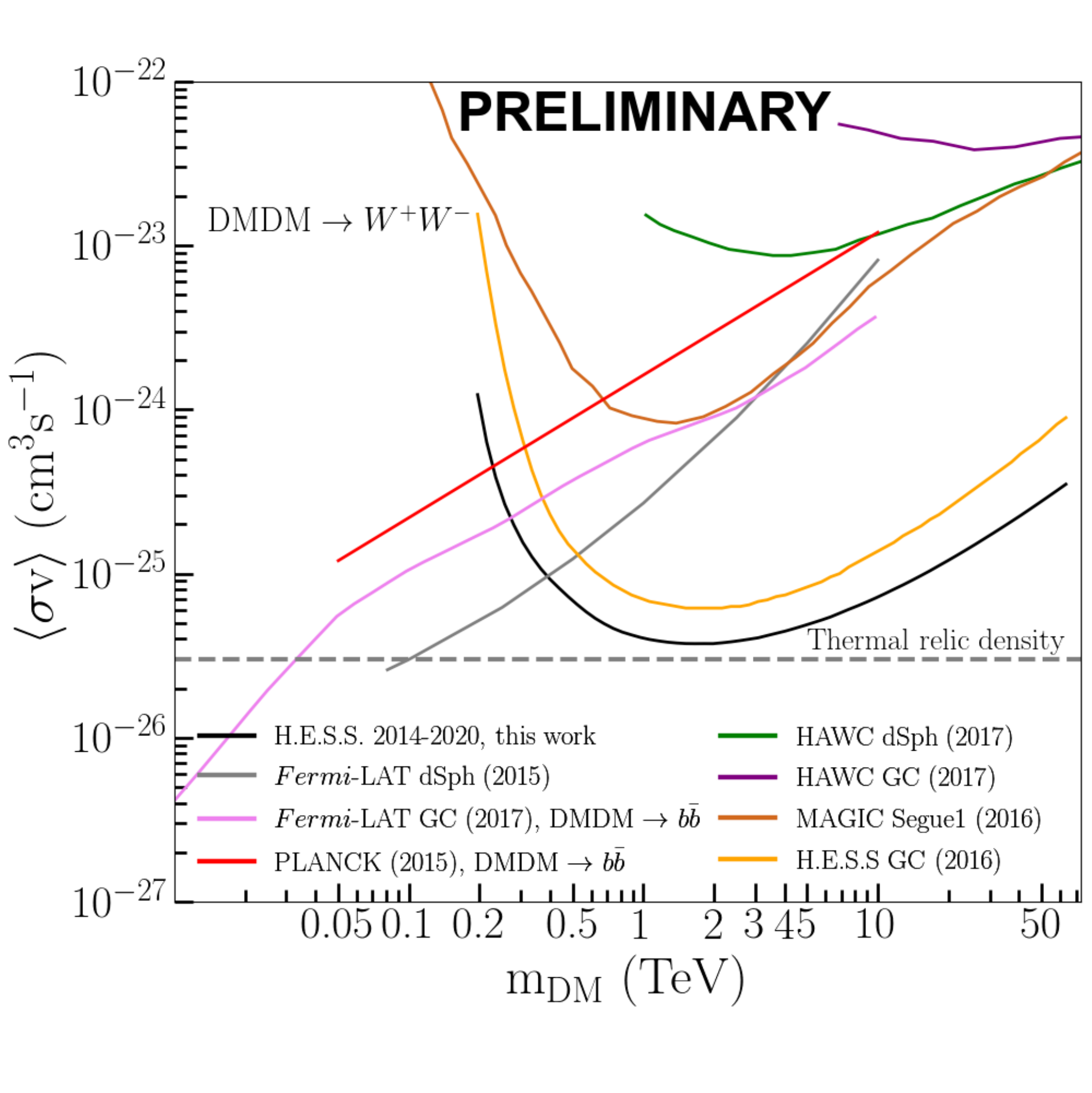}
\caption{{\it Left panel:} 95\% C. L. upper limits on the velocity-weighted annihilation cross section $\langle \sigma v \rangle$ as a function of the DM mass $m_{\rm DM}$ for the  W$^+$W$^-$ channel. The upper limits are derived from the H.E.S.S. observations taken from 2014 to 2020 and include the systematic uncertainty. The black solid line shows the observed limits. The black dashed line shows the mean expected limits. The green and yellow bands show the 68\% and 95\% C.L. statistical containment bands, respectively. The value of the natural scale expected for the thermal-relic WIMPs is shown as the horizontal grey long-dashed line. {\it Right panel:} Comparison of the current upper limits on the W$^+$W$^-$ channels with other experiments. The limits obtained with 254 hours of observations of the GC with H.E.S.S.~\cite{Abdallah:2016ygi} (orange line), the observation of the GC with HAWC~\cite{Abeysekara_2018} (purple line), the observations of 15 dwarf galaxy satellites of the Milky Way~\cite{Ackermann:2015zua} (grey line) and the observation of the GC~\cite{Ackermann_2017} with the Fermi satellite (here $b\bar{b}$ channel, pink line), 158 hours of observations of the dwarf galaxy Segue 1 with MAGIC~\cite{Ahnen:2016qkx} (brown line), the combined analysis of observations of 15 dwarf galaxies by HAWC~\cite{Albert_2018} (green line) and the cosmic wave background with PLANCK~\cite{Adam:2015rua} (here $b\bar{b}$ channel, red line) are shown.}
\label{fig:results}
\end{figure*}

Neither this analysis nor a cross-check one result in any significant gamma-ray excess in the ON with respect to the OFF regions. 
We derive 95\% C.L. upper limits on $\langle \sigma v \rangle$, exploring the self-annihilation of WIMPs with masses from 200 GeV up to 70 TeV. Fig.~\ref{fig:results} shows the 95\% C.L. upper limits in the gauge boson $W^+W^-$ channels for the Einasto profile with $\rho_s=0.079$ GeVcm$^{-3}$, r$_{s}=$20 kpc and $\alpha_{s}$=0.17. To obtain the expected limits, 300 Poisson realizations of the background OFF regions are produced. The mean expected upper limits and the 68\% and 95\% containment bands are extracted from the mean, the 1 and 2$\sigma$ standard deviation of the distribution of log10($\langle \sigma v \rangle$) computed from the realizations. Mean expected limits and containment bands are plotted in Fig.~\ref{fig:results}. The 95\% C.L. observed limits reach 3.7$\times 10^{-26}$cm$^3$s$^{-1}$ for a DM particle mass of 1.5 TeV in the $W^+W^-$ channel. The limits are close to the $\langle \sigma v \rangle$ values expected for DM particle annihilating with the thermal-relic cross section.
Due to longer observational live time, the statistics of the observations taken for the IGS program is larger than the one used in the previous H.E.S.S. results~\cite{Abdallah:2016ygi}. In addition, the IGS observations provide higher sensitivity because were carried out with the full CT1-5 array of H.E.S.S.. At 1.5 TeV DM mass, an improvement factor of 1.6 with respect to the results shown in Ref.~\cite{Abdallah:2016ygi} is obtained. 

In the right panel of Fig.~\ref{fig:results}, limits on $\langle \sigma v \rangle$ from different experiment are compiled. 
The current gamma-ray limits together with the previous H.E.S.S. limits from 254 hours of observation of the Galactic Center~\cite{Abdallah:2016ygi}, the limits from the observation of 15 dwarf galaxy satellites of the Milky Way~\cite{Ackermann:2015zua} and from the observation of the GC with $Fermi$-LAT~\cite{Ackermann_2017}, the HAWC limits from the observation of the GC~\cite{Abeysekara_2018} as well as the combined analysis of the observation of 15 dwarf galaxies~\cite{Albert_2018}, the limits from 158 hours of observation of the dwarf galaxy Segue 1 with MAGIC~\cite{Ahnen:2016qkx} and the limits from the cosmic microwave background with PLANCK~\cite{Adam:2015rua} are shown.
For the  W$^+$W$^-$ channel and particle masses above $\sim$300~GeV, the present H.E.S.S. constraints surpass the {\it Fermi}-LAT limits. VHE observations of the central region of the Milky Way with  H.E.S.S. contribute uniquely to the study of the WIMP paradigm for TeV DM. Multi-TeV DM models like the benchmark candidates Wino and Higgsino (see, for instance, Ref.~\cite{Rinchiuso:2020skh} and references therein), which are predicted in simple extensions to the Standard Model, can be explored with the unprecedented dataset of the IGS observations. The IGS precedes the Southern-site sky observations with CTA~\cite{Moulin:2019oyc} and is one of the important legacies of H.E.S.S..

\section{Acknowledgements}
Full H.E.S.S. acknowledgements can be found \href{https://www.mpi-hd.mpg.de/hfm/HESS/pages/publications/auxiliary/HESS-Acknowledgements-2021.html}{here}.

\section{Full author list}
H.~Abdalla$^{1}$, 
F.~Aharonian$^{2,3,4}$, 
F.~Ait~Benkhali$^{3}$, 
E.O.~Ang\"uner$^{5}$, 
C.~Arcaro$^{6}$, 
C.~Armand$^{7}$, 
T.~Armstrong$^{8}$, 
H.~Ashkar$^{9}$, 
M.~Backes$^{1,6}$, 
V.~Baghmanyan$^{10}$
% ...

\medskip

\noindent
$^{1}$University of Namibia, Department of Physics, Private Bag 13301, Windhoek 10005, Namibia\\
$^{2}$Dublin Institute for Advanced Studies, 31 Fitzwilliam Place, Dublin 2, Ireland\\
$^{3}$Max-Planck-Institut f\"ur Kernphysik, P.O. Box 103980, D 69029 Heidelberg, Germany\\
$^{4}$High Energy Astrophysics Laboratory, RAU,  123 Hovsep Emin St  Yerevan 0051, Armenia\\
$^{5}$Aix Marseille Universit\'e, CNRS/IN2P3, CPPM, Marseille, France\\
$^{6}$Centre for Space Research, North-West University, Potchefstroom 2520, South Africa\\
$^{7}$Laboratoire d'Annecy de Physique des Particules, Univ. Grenoble Alpes, Univ. Savoie Mont Blanc, CNRS, LAPP, 74000 Annecy, France\\
$^{8}$University of Oxford, Department of Physics, Denys Wilkinson Building, Keble Road, Oxford OX1 3RH, UK\\
$^{9}$IRFU, CEA, Universit\'e Paris-Saclay, F-91191 Gif-sur-Yvette, France\\
$^{10}$Instytut Fizyki J\c{a}drowej PAN, ul. Radzikowskiego 152, 31-342 Krak{\'o}w, Poland\\
% ...

%
%
%


\begin{thebibliography}{99}
\bibitem{Adam:2015rua}
R.~Adam \textit{et al.} [Planck],
{\it Planck 2015 results. I. Overview of products and scientific results},
Astron. Astrophys. \textbf{594} (2016)
\bibitem{Bertone:2010zza}
Bertone, G. (Ed.). (2010). {\it Particle Dark Matter: Observations, Models and Searches.} Cambridge: Cambridge University Press.
\bibitem{Feng:2010gw}
J.~L.~Feng, {\it Dark Matter Candidates from Particle Physics and Methods of Detection}, Ann. Rev. Astron. Astrophys. \textbf{48} (2010)
\bibitem{Roszkowski:2017nbc}
L.~Roszkowski, E.~M.~Sessolo and S.~Trojanowski, {\it WIMP dark matter candidates and searches\textemdash{}current status and future prospects}, Rept. Prog. Phys. \textbf{81} (2018) no.6
\bibitem{Jungman:1995df}
G.~Jungman, M.~Kamionkowski and K.~Griest, {\it Supersymmetric dark matter}, Phys. Rept. \textbf{267} (1996)
\bibitem{Kahlhoefer:2017dnp}
F.~Kahlhoefer, {\it Review of LHC Dark Matter Searches}, Int. J. Mod. Phys. A \textbf{32} (2017) no.13, 
\bibitem{Schumann:2019eaa}
M.~Schumann, {\it Direct Detection of WIMP Dark Matter: Concepts and Status},
J. Phys. G \textbf{46} (2019) no.10
\bibitem{Strigari:2018utn}
L.~E.~Strigari, {\it Dark matter in dwarf spheroidal galaxies and indirect detection: a review},
Rept. Prog. Phys. \textbf{81} (2018) no.5
\bibitem{Ghez:2008ms}
A.~M.~Ghez, S.~Salim, N.~N.~Weinberg, J.~R.~Lu, T.~Do, J.~K.~Dunn, K.~Matthews, M.~Morris, S.~Yelda and E.~E.~Becklin, \textit{et al.}, {\it Measuring Distance and Properties of the Milky Way's Central Supermassive Black Hole with Stellar Orbits},
Astrophys. J. \textbf{689} (2008)
\bibitem{Springel:2008by}
V.~Springel, S.~D.~M.~White, C.~S.~Frenk, J.~F.~Navarro, A.~Jenkins, M.~Vogelsberger, J.~Wang, A.~Ludlow and A.~Helmi,
{\it A blueprint for detecting supersymmetric dark matter in the Galactic halo} 
\bibitem{Navarro:1996gj}
J.~F.~Navarro, C.~S.~Frenk and S.~D.~M.~White, {\it A Universal density profile from hierarchical clustering},
Astrophys. J. \textbf{490} (1997)
\bibitem{Abdallah:2018qtu}
H.~Abdallah \textit{et al.} [HESS], {\it Search for $\gamma$-Ray Line Signals from Dark Matter Annihilations in the Inner Galactic Halo from 10 Years of Observations with H.E.S.S.},
Phys. Rev. Lett. \textbf{120} (2018) no.20
\bibitem{Abdallah:2016ygi}
H.~Abdallah \textit{et al.} [H.E.S.S.], {\it Search for dark matter annihilations towards the inner Galactic halo from 10 years of observations with H.E.S.S},
Phys. Rev. Lett. \textbf{117} (2016) no.11
\bibitem{Aharonian:2006pe}
F.~Aharonian \textit{et al.} [H.E.S.S.], {\it Observations of the Crab Nebula with H.E.S.S},
Astron. Astrophys. \textbf{457} (2006)
\bibitem{2009APh32231D}
de Naurois, M. \& Rolland, L., {\it A high performance likelihood reconstruction of {$\gamma$}-rays for imaging atmospheric Cherenkov telescopes}, Astroparticle Physics, 32, 231.
\bibitem{Aharonian:2009zk} 
Aharonian, F. {\it et al.} [H.E.S.S.], {\it Spectrum and variability of the Galactic Center VHE gamma-ray source HESS J1745-290}, A\&A, 503, 817. 
\bibitem{Abramowski:2016mir}
A.~Abramowski \textit{et al.} [H.E.S.S.], {\it Acceleration of petaelectronvolt protons in the Galactic Centre},
Nature \textbf{531} (2016)
\bibitem{H.E.S.S.:2018zkf}
H.~Abdalla \textit{et al.} [H.E.S.S.], {\it The H.E.S.S. Galactic plane survey}, Astron. Astrophys. \textbf{612} (2018), A1
\bibitem{Cirelli:2010xx}
M.~Cirelli, G.~Corcella, A.~Hektor, G.~Hutsi, M.~Kadastik, P.~Panci, M.~Raidal, F.~Sala and A.~Strumia, {\it PPPC 4 DM ID: A Poor Particle Physicist Cookbook for Dark Matter Indirect Detection}, JCAP \textbf{03} (2011)
\bibitem{Silverwood:2014yza}
H.~Silverwood, C.~Weniger, P.~Scott and G.~Bertone, {\it A realistic assessment of the CTA sensitivity to dark matter annihilation},
JCAP \textbf{03} (2015)
\bibitem{Lefranc:2015pza}
V.~Lefranc, E.~Moulin, P.~Panci and J.~Silk, {\it Prospects for Annihilating Dark Matter in the inner Galactic halo by the Cherenkov Telescope Array}, Phys. Rev. D \textbf{91} (2015) no.12
\bibitem{Moulin:2019oyc}
E. Moulin, J. Carr, J. Gaskins, M. Doro, C. Farnier, M. Wood and H. Zechlin, {\it Science with the Cherenkov Telescope Array: Dark Matter Programme}, World Scientific,
\bibitem{Cowan:2010js}
G.~Cowan, K.~Cranmer, E.~Gross and O.~Vitells,
{\it Asymptotic formulae for likelihood-based tests of new physics},
Eur. Phys. J. C \textbf{71} (2011)
\bibitem{Parsons:2014voa}
R.~D.~Parsons and J.~A.~Hinton, {\it A Monte Carlo Template based analysis for Air-Cherenkov Arrays}, Astropart. Phys. \textbf{56} (2014)
\bibitem{Bertone:2004pz}
G.~Bertone, D.~Hooper and J.~Silk, {\it Particle dark matter: Evidence, candidates and constraints}, Phys. Rept. \textbf{405} (2005)
\bibitem{Abeysekara_2018} 
Abeysekara, A.~U., Albert, A., Alfaro, R., {\it et al.}, 
{\it A Search for Dark Matter in the Galactic Halo with HAWC},
2018, JCAP, 2018, 049. 
\bibitem{Rinchiuso:2020skh}
L.~Rinchiuso, O.~Macias, E.~Moulin, N.~L.~Rodd and T.~R.~Slatyer, {\it Prospects for detecting heavy WIMP dark matter with the Cherenkov Telescope Array: The Wino and Higgsino}, Phys. Rev. D \textbf{103} (2021) no.2
\bibitem{Ackermann:2015zua}
M.~Ackermann \textit{et al.} [Fermi-LAT], {\it Searching for Dark Matter Annihilation from Milky Way Dwarf Spheroidal Galaxies with Six Years of Fermi Large Area Telescope Data}, Phys. Rev. Lett. \textbf{115} (2015) no.23
\bibitem{Albert_2018} 
Albert, A., Alfaro, R., Alvarez, C., {\it et al.} 2018, Astrophys. J., 853, 154. 
\bibitem{Ahnen:2016qkx}
M.~L.~Ahnen \textit{et al.} [MAGIC and Fermi-LAT], {\it Limits to Dark Matter Annihilation Cross-Section from a Combined Analysis of MAGIC and Fermi-LAT Observations of Dwarf Satellite Galaxies},
JCAP \textbf{02} (2016)
\bibitem{Ackermann_2017}
M.~Ackermann \textit{et al.} [Fermi-LAT], {\it The Fermi Galactic Center GeV Excess and Implications for Dark Matter},
Astrophys. J. \textbf{840} (2017) no.1
\bibitem{Abdallah:2021czg}
H.~Abdalla \textit{et al.} [H.E.S.S.], {\it Search for dark matter annihilation signals from unidentified Fermi-LAT objects with H.E.S.S.}, [arXiv:2106.00551 [astro-ph.HE]].
\bibitem{Abdallah:2021kzs}
H.~Abdallah \textit{et al.} [H.E.S.S.], {\it Search for dark matter annihilation in the Wolf-Lundmark-Melotte dwarf irregular galaxy with H.E.S.S.},
Phys. Rev. D \textbf{103} (2021) no.10
\bibitem{Abdallah:2020sas}
H.~Abdallah \textit{et al.} [H.E.S.S.], {\it Search for dark matter signals towards a selection of recently detected DES dwarf galaxy satellites of the Milky Way with H.E.S.S.}, Phys. Rev. D \textbf{102} (2020) no.6

\end{thebibliography}
\end{document}